\documentclass[aps,prd,preprint,superscriptaddress,tightenlines,nofootinbib]{revtex4}

\pdfoutput=1

\usepackage{graphicx}
\usepackage{dcolumn}
\usepackage{bm}

\newcommand{\etal}{{\it et al.}}

\begin{document}

\preprint{CLNS 06/1982}       
\preprint{CLEO 06-22}         

\title{A Study of Exclusive Charmless Semileptonic  $B$ Decay and $|V_{ub}|$}

\author{N.~E.~Adam}
\author{J.~P.~Alexander}
\author{K.~Berkelman}
\author{D.~G.~Cassel}
\author{J.~E.~Duboscq}
\author{R.~Ehrlich}
\author{L.~Fields}
\author{L.~Gibbons}
\author{R.~Gray}
\author{S.~W.~Gray}
\author{D.~L.~Hartill}
\author{B.~K.~Heltsley}
\author{D.~Hertz}
\author{C.~D.~Jones}
\author{J.~Kandaswamy}
\author{D.~L.~Kreinick}
\author{V.~E.~Kuznetsov}
\author{H.~Mahlke-Kr\"uger}
\author{T.~O.~Meyer}
\author{P.~U.~E.~Onyisi}
\author{J.~R.~Patterson}
\author{D.~Peterson}
\author{J.~Pivarski}
\author{D.~Riley}
\author{A.~Ryd}
\author{A.~J.~Sadoff}
\author{H.~Schwarthoff}
\author{X.~Shi}
\author{S.~Stroiney}
\author{W.~M.~Sun}
\author{T.~Wilksen}
\author{M.~Weinberger}
\affiliation{Cornell University, Ithaca, New York 14853}
% ...
\collaboration{CLEO Collaboration} 
\noaffiliation


\date{March 23, 2007}

\begin{abstract} 
We study semileptonic $B$ decay to the exclusive charmless  states 
$\pi$, $\rho/\omega$, $\eta$ and $\eta^\prime$  using the 16~fb$^{-1}$
CLEO $\Upsilon(4S)$ data sample.  
We find 
 ${\cal B}(B^0 \to \pi^-\ell^+\nu)=(1.37 \pm 0.15_{\text{stat}} \pm 0.11_{\text{sys}} ) \times 10^{-4}$
and ${\cal B}(B^0 \to \rho^- \ell^+\nu)=(2.93 \pm 0.37_{\text{stat}} \pm 0.37_{\text{sys}})\times 10^{-4}$,
and find evidence  for $B^+\to\eta^{\prime}\ell^+\nu$, with 
${\cal B}(B^+\to\eta^{\prime}\ell^+\nu)=(2.66\pm0.80_{\text{stat}}\pm0.56_{\text{sys}})\times 10^{-4}$.
From our  $B\to\pi\ell\nu$ rate for $q^2>16$ GeV$^{2}$ and lattice QCD
we find  
$|V_{ub}| = (3.6\pm 0.4_{\text{stat}} \pm 0.2_{\text{syst}} {}^{+0.6}_{-0.4\text{thy}})\times 10^{-3}$.
\end{abstract}

\maketitle

The magnitude of the Cabibbo-Kobayashi-Maskawa (CKM) matrix \cite{bb:CKM} element $V_{ub}$ plays a central role in tests of the matrix's unitarity, which is the main thrust of the current heavy-flavor program.  Violation of unitarity would signal existence of new classes of fundamental particles or forces.  Robust determination of $|V_{ub}|$ has been the subject of considerable theoretical and experimental effort for well over a decade, and remains one of the highest priorities of flavor physics, yet it remains one of the most uncertain parameters of
the matrix.  A recent determination, dominated by inclusive measurements, has marginal agreement with other inputs to the unitarity constraints \cite{Bona:2006ah}.
 Exclusive charmless semileptonic decays, particularly $B\to \pi\ell\nu$, provide an alternate route to determination of $|V_{ub}|$.

This letter presents a study of the decays $B\to \pi\ell\nu$, $B\to\rho\ell\nu$ and 
$B\to\eta^{(\prime)}\ell\nu$ based on the full 15.4 million $B\bar{B}$ data sample collected at the 
$\Upsilon(4S)$ with
the CLEO II, II.5, and III detectors~\cite{Kubota:1992ww,Hill:1998ea,Peterson:1994cd} at
the Cornell Electron Storage Ring (CESR).  These results include an important crosscheck of the leading measurement of the critical $B^0\to\pi^-\ell^+\nu$ branching fraction~\cite{bb:BABARy2007} using a different analysis technique in a symmetric beam environment.  We also present the most precise study of $B\to\rho\ell\nu$,  which is an important background process to $B\to\pi\ell\nu$.  The results include the first partial rate measurement for a region of $B\to\rho\ell\nu$ phase space (helicity angle over $90^\circ$) that
provides about half of the total $B\to\rho\ell\nu$ background in the $B\to\pi\ell\nu$ signal region.  This benefits both this and other $B\to\pi\ell\nu$ measurements that have a background from $B\to\rho\ell\nu$.
Finally, we find first evidence for the decay $B\to\eta^\prime\ell\nu$.  This channel is predicted to be enhanced relative to $B\to\eta\ell\nu$ through coupling to the singlet component of the $\eta^\prime$~\cite{etap:CSK,etap:Neubert2003}, which could resolve the unexpectedly large decay rate for $B\to\eta^\prime K$.  These results supersede those of references~\cite{Athar:2003yg,Alexander:1996qu}.

Hadronic form factors (FFs) present  challenges experimentally and theoretically for determination of $|V_{ub}|$.  For the decay $B\to V_u\ell\nu$, where $V_u$ is a charmless vector meson, the partial width is
\begin{equation}
\frac{d\Gamma}{dq^2\,dC_\theta} =\kappa kq^2 \left[ S^2_\theta H_0^2 +
\frac{C_-^2H_+^2 +  
C_+^2H_-^2}{2} \right].
\end{equation}
Here, $\kappa= \frac{|V_{ub}|^2 G_F^2}{128\pi^3M_B^2}$,  $k$ is the $V_u$ momentum, $C_\theta$ ($S_\theta$) is the cosine (sine) of the angle $\theta_{W\ell}$ between the charged lepton in the virtual $W$ ($W^*$) rest frame and the $W^*$ in the $B$ rest frame, and $C_\pm=1\pm C_\theta$.  
$H_\pm$ and $H_0$ are the magnitudes of the $W$ helicity amplitudes, which can be expressed in the massless lepton limit in terms of three $q^2$--dependent FFs~\cite{Gilman_and_Singleton}.
For a final state pseudoscalar meson $P_u$, $H_{\pm}=0$, and the rate depends on a single FF 
$f_+(q^2)$:
\begin{equation}
\frac{d\Gamma\left(B\to P_u\ell\nu\right)}{dq^2}=\left|V_{ub}\right|^2\frac{G_F^2}{24\pi^3}k^3\left|f_+\left(q^2\right)\right|^2.
\label{eq:dgammadq2}
\end{equation}

Experimental measurement of the branching fractions requires knowledge of the $q^2$-dependence of the FFs. All current measurements follow the previous CLEO analysis \cite{Alexander:1996qu} and mitigate FF shape uncertainties by measuring partial rates in regions of phase space.
Subsequent determination of $|V_{ub}|$ requires theoretical input on the FF normalization. Recent calculation of $f_{+}(q^2)$ from lattice QCD using dynamical quarks~\cite{Gulez:2006dt} represent a marked theoretical advance. 

The analysis uses the missing four-momentum, $p_{\text{miss}}\equiv p_{\text{CM}}- p_{\text{visible}}$ to estimate the neutrino four-momentum, $p_\nu$.  All CLEO detector configurations~\cite{Kubota:1992ww,Hill:1998ea,Peterson:1994cd} provide acceptance over more than 90\% of the full $4\pi$ solid angle for charged particles (momentum resolution of 0.6\% at 2 GeV$/c$) and for photons (average $\pi^0$ mass resolution of 6 MeV/$c^2$).  We utilize global event reconstruction and particle identification algorithms, optimized for neutrino reconstruction, outlined in detail elsewhere~\cite{PRD,waprl}.

Signal leptons are identified with over 90\% efficiency in the range $1.0 < p_\ell < 3.0 \text{ GeV}/c$.  We reconstruct hadronic $X_u$ candidates $\pi^0$, $\pi^\pm$, $\eta$, $\rho^0$, $\rho^\pm$, $\omega$, and $\eta^\prime$.  We require the reconstructed $\pi^0$ ($\eta$) mass in the $\gamma\gamma$ mode to be within two standard deviations, about 8 (26) MeV/$c^2$, of the nominal $\pi^0$ ($\eta$) mass.  These candidates are kinematically fit to improve the momentum resolution.  Candidate $\eta$ ($\omega$) decays with $\pi^+\pi^-\pi^0$ invariant mass within 10 (30) MeV/$c^2$ of the nominal $\eta$ ($\omega$) mass and $\rho$ candidates with $\pi\pi$ invariant mass within 285 MeV/$c^2$ of the $\rho$ mass are accepted.  The decay of the $\eta^\prime$ is reconstructed in both the $\pi\pi\gamma$ and $\eta\pi^+\pi^-$ final states, with $\eta\to\gamma\gamma$. For $\eta^{\prime}\to\pi\pi\gamma$, the invariant mass of the reconstructed $\eta^{\prime}$ must be within 2.5 (2.75) standard deviations of the nominal $\eta^{\prime}$ mass for $q^2<10$ ($q^2 > 10$) GeV$^2$.  For $\eta^{\prime}\to\eta\pi\pi$ the quadrature sum of the number of standard deviations from the $\eta$ and $\eta^{\prime}$ masses must be less than 3.75. 

In events with multiple undetected particles,  $p_{\text{miss}}$ represents $p_\nu$ poorly, causing $B\to X_c\ell\nu$ decays to smear into the signal region of (much rarer) $B\to X_u\ell\nu$ processes, while smearing much of the $X_u\ell\nu$ out.  This mechanism provides our dominant background contribution.  Therefore, we reject events with multiple identified leptons, which are usually accompanied by multiple neutrinos, and events with net charge $|\Delta Q| > 0$, an indication of missed particles.  In the pseudoscalar decay modes, we do achieve some additional precision by independently considering the more contaminated $|\Delta Q| = 1$ sample.  We further require that $M^2_{\text{miss}}$ be consistent with a massless neutrino within experimental resolution.  Because our resolution on $E_{\text{miss}}$, $\sigma_{E_{\text{miss}}}$ ($\approx 0.2$ GeV), 
is larger than that for $\vec{p}_{\text{miss}}$ ($\approx 0.1$ GeV), 
the $M^2_{\text{miss}}$ resolution scales as $\sim 2E_{\text{miss}}/\sigma_{E_{\text{miss}}}$, so we require $\left|M^2_{\text{miss}}/2E_{\text{miss}}\right| \lesssim 0.5$ GeV -- a requirement that is approximately constant in $E_{\text{miss}}$ resolution.  The criterion was optimized, using only independent Monte Carlo samples, for maximum signal significance.

Candidate $B$ decays are formed by combining the signal lepton and reconstructed neutrino with $X_u$ candidates, taking $p_\nu = \left( \left|\vec{p}_{\text{miss}}\right|,\vec{p}_{\text{miss}}\right)$ for  reconstruction. Energy and momentum conservation requirements are cast in terms of $\Delta E \equiv E_{X_u} + E_{\ell} + E_\nu - E_{\text{beam}}$ and $M_{h\ell\nu}\equiv\sqrt{E_{\text{beam}}^2-\left|\vec{p}_{X_u}+\vec{p}_\ell + \alpha\vec{p}_\nu\right|^2}$, which peak at zero and $M_B$ for signal decays.  Because the neutrino resolution dominates the $\Delta E$ resolution, we improve the $M_{h\ell\nu}$ resolution by choosing $\alpha$ such that $E_{X_u} + E_\ell + \alpha E_\nu - E_{\text{beam}} = 0$. Before computing the kinematic variables $q^2$ and $\cos\theta_{W\ell}$, the reconstructed neutrino is rotated to force $M_{h\ell\nu}$ to the value of $M_B$ to optimize resolution in these variables.  This procedure produces resolutions of 0.3 GeV$^2$ and 0.03 for $q^2$ and $\cos\theta_{W\ell},$ respectively.

Continuum $e^+e^-\to q\bar{q}$, where $q=u,~d,~s,~\text{or}~c,$ can be distinguished from $B\bar{B}$ events by examining a variety of event shape variables.  Specifically, we examine the ratio of the second to zeroth Fox-Wolfram moments~\cite{Fox:1978vu}, the angle between the candidate thrust axis and the thrust axis of the rest of the event, the angle between the event thrust axis and the beam, and the momentum flow into nine double cones whose axes are centered on the beam.  A Fisher discriminant~\cite{Fisher} is computed using these twelve variables. Requirements on this discriminant are tuned for each decay mode and each $q^2$ and $\cos\theta_{W\ell}$ region to maximize signal significance.  Due to their jet-like nature, continuum backgrounds typically reconstruct with low $q^2$ and can be largely isolated at $q^2 < 2$ GeV$^2$.

We fit seven coarse bins~\cite{PRD} of the $\Delta E$ versus $M_{h\ell\nu}$ distributions, separately reconstructed within each phase space region summarized in Table~\ref{tab:results} to allow extraction of a partial branching fraction for each region.  The bins span  $5.1750~\text{GeV}/c^2<M_{h\ell\nu}<5.2875~\text{GeV}/c^2$ and $-0.75~\text{GeV}< \Delta E < 0.25~\text{GeV}$.  We are primarily sensitive to signal within the bin defined by $5.2650~\text{GeV}/c^2<M_{h\ell\nu}<5.2875~\text{GeV}/c^2$ and $-0.15~\text{GeV} < \Delta E < 0.25~\text{GeV}$.  We also use coarse lineshape information to help separate signals involving resonances in the final state from background. For the $\rho\ell\nu$ ($\omega\ell\nu$) mode, we accomplish this by separation of the reconstructed distributions into three 195 MeV/$c^2$ (20 MeV/$c^2$) intervals in reconstructed ${\pi\pi}$  (${3\pi}$)  mass centered on the nominal resonance mass.  In $\eta^\prime\to\pi^+\pi^-\gamma$, we use four $M_{\pi\pi}$ bins which span the range from 300 MeV/$c^2$ to 900 MeV/$c^2$. The fit utilizes a binned maximum likelihood approach extended to include the finite statistics  of the fit components~\cite{Barlow_and_Beeston}. 

A GEANT-based Monte Carlo~\cite{GEANT} is used to model the distributions for the signal, $b\to c$ background, and backgrounds from other $B\to X_u\ell\nu$ decays that are not explicitly fit, {\it e.g.} $B\to a_0\ell\nu$.  The signal Monte Carlo (MC) is divided at the generator level into the phase space regions of Table~\ref{tab:results}.  We obtain a full set of reconstructed distributions from each which are normalized in the fit to a rate parameter for that region.  We fit all modes (and their subregions) simultaneously.  The data in one phase space region thereby controls the normalization for that region's cross-feed into all other regions and modes, minimizing our dependence on {\em a priori} form factor and branching fraction information. The input form factors for the signal MC samples are derived from an unquenched lattice (LQCD) calculation~\cite{Gulez:2006dt}  for $\pi\ell\nu$  and a  light cone sum rules (LCSR) calculation~\cite{ball04rho} for $\rho/\omega\ell\nu$.   Data taken below the $\Upsilon(4S)$, smoothed to reduce statistical fluctuations~\cite{PRD}, is used to model residual continuum backgrounds.  Small contributions from fake signal leptons are modeled using purely hadronic data and measured lepton fake rates.  

To increase sensitivity to $\eta^\prime\ell\nu$ decay, the candidates are divided into two bins in $q^2$: one greater than 10~GeV$^2$, which typically has higher $B\to X_c\ell\nu$ background, and one less than 10~GeV$^2$.  This division enhances the the significance of the expected signal-rich region without introducing form factor dependence associated with a strict $q^2$ requirement.  We fit only for the total $B\to\eta^\prime\ell\nu$ rate by using the ISGW2 model~\cite{isgw2} to fix the ratio of efficiency corrected yields in these two regions.

We minimize multiple candidates per event to simplify statistical interpretation.  Within each $\pi$,  $\eta$ or $\eta^\prime\to\eta\pi^+\pi^-$ mode, we allow only one candidate within the fit's $\Delta E-M_{h\ell\nu}$ region.  Within a mode, multiple candidates in that region are resolved by choosing that with the smallest $|\Delta E|$. For the vector and the $\eta^\prime\to\pi^+\pi^-\gamma$ modes, the $|\Delta E|$ criterion combined with large combinatorics outside the signal region induces efficiency loss.  To mitigate loss, we select the best candidate separately within each $M_{\pi\pi}$ or $M_{3\pi}$ range.  This procedure induces only very slight peaking in backgrounds, which is modeled by the MC generated distributions used to fit the data.  While a given event can still contribute to more than one mode (or mass range), tests of the fitter using numerous MC samples as toy data sets verify that signal yields and statistical errors are extracted without bias.

 Signal selection efficiencies, averaged over phase space, for the restricted signal region range from $\approx 4\%$ for the $\pi^\pm\ell\nu$ mode down to $\approx 0.5\%$ for the $\eta^\prime\ell\nu$ mode.

The nominal fit includes one free parameter for the yield in each of the 4 (5) regions in $\pi\ell\nu$ ($\rho\ell\nu$) phase space and free parameters for the total $\eta\ell\nu$ and $\eta^\prime\ell\nu$ yields.  Within each phase space region, the charged and neutral $\pi$ and $\rho$ rates were fixed to be consistent with expectations from isospin, $B$ lifetime differences, and relative charged/neutral $B$ production at the $\Upsilon(4S)$~\cite{Alexander:2005cx}.  The total $B^+\to\omega\ell^+\nu$ rate is constrained to the $B^+\to\rho^0\ell^+\nu$ rate.  To minimize systematic biases, the normalization of $b\to c$ background floats freely for each mode and for each $|\Delta Q|$ sample.  The sum of the signal rates and the feed-down from higher mass $B\to X_u\ell\nu$ rates is constrained by recent inclusive $b\to u$ measurements at the lepton endpoint~\cite{babarinclep}.  There are a total of 26 parameters.

\begin{figure}[t]
\includegraphics[scale=0.4]{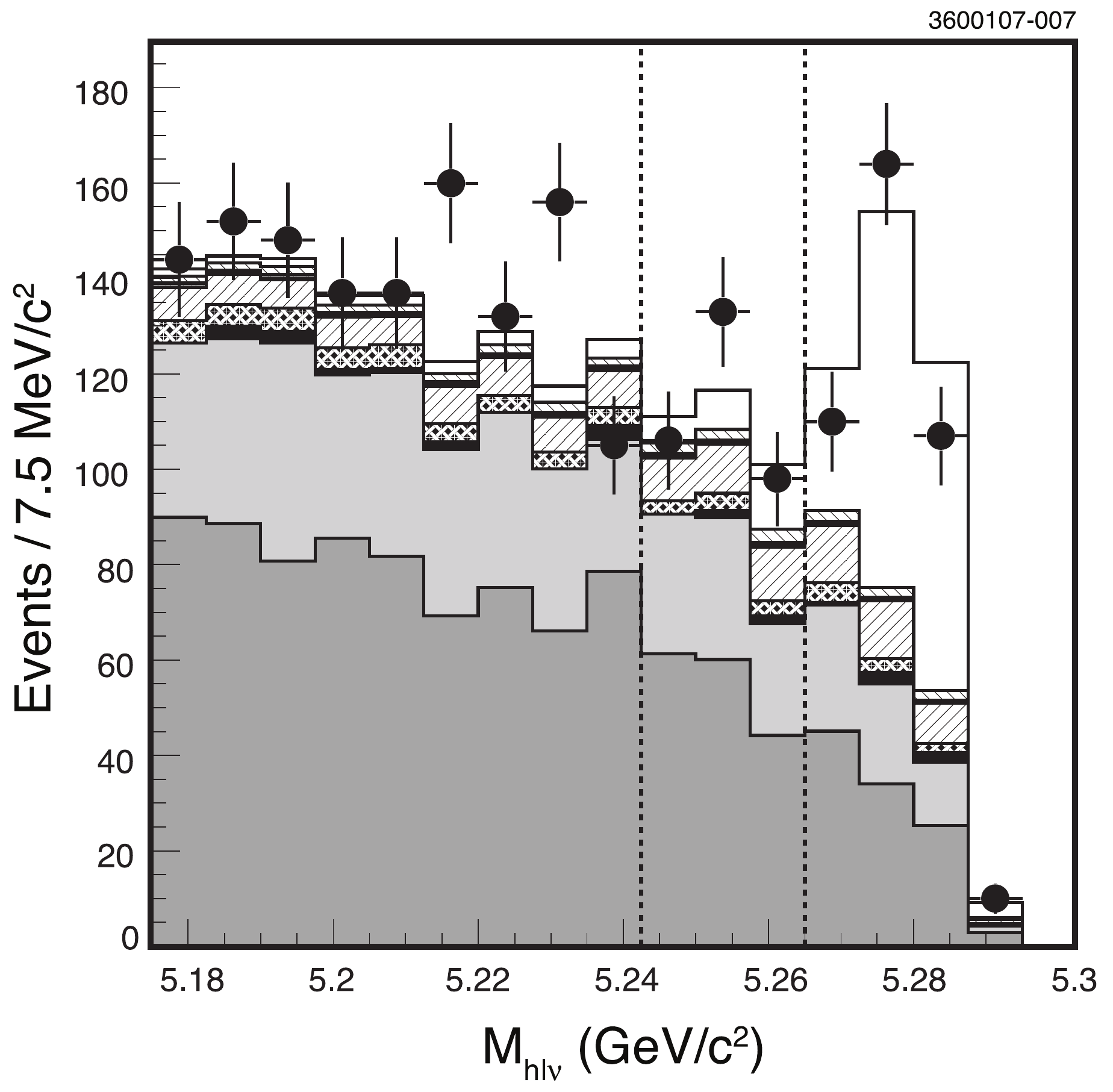}
\caption{$M_{h\ell\nu}$ for candidates in the $\Delta E$ signal region [$-0.15~\text{GeV}, 0.25~\text{GeV}$] for the $B\to\pi\ell\nu$ modes combined and summed over $q^2$.  Dotted lines indicate the binning used by the fitter.  Fit components, from bottom to top are, $b\to c$, continuum, fake signal leptons, other $B\to X_u\ell\nu$, $B\to\eta\ell\nu$ cross-feed, $B\to\rho\ell\nu$ cross-feed, $B\to\pi\ell\nu$ cross-feed, and signal.}
\label{fig:pifit}
\end{figure}

\begin{figure}[t]
\includegraphics[scale=0.4]{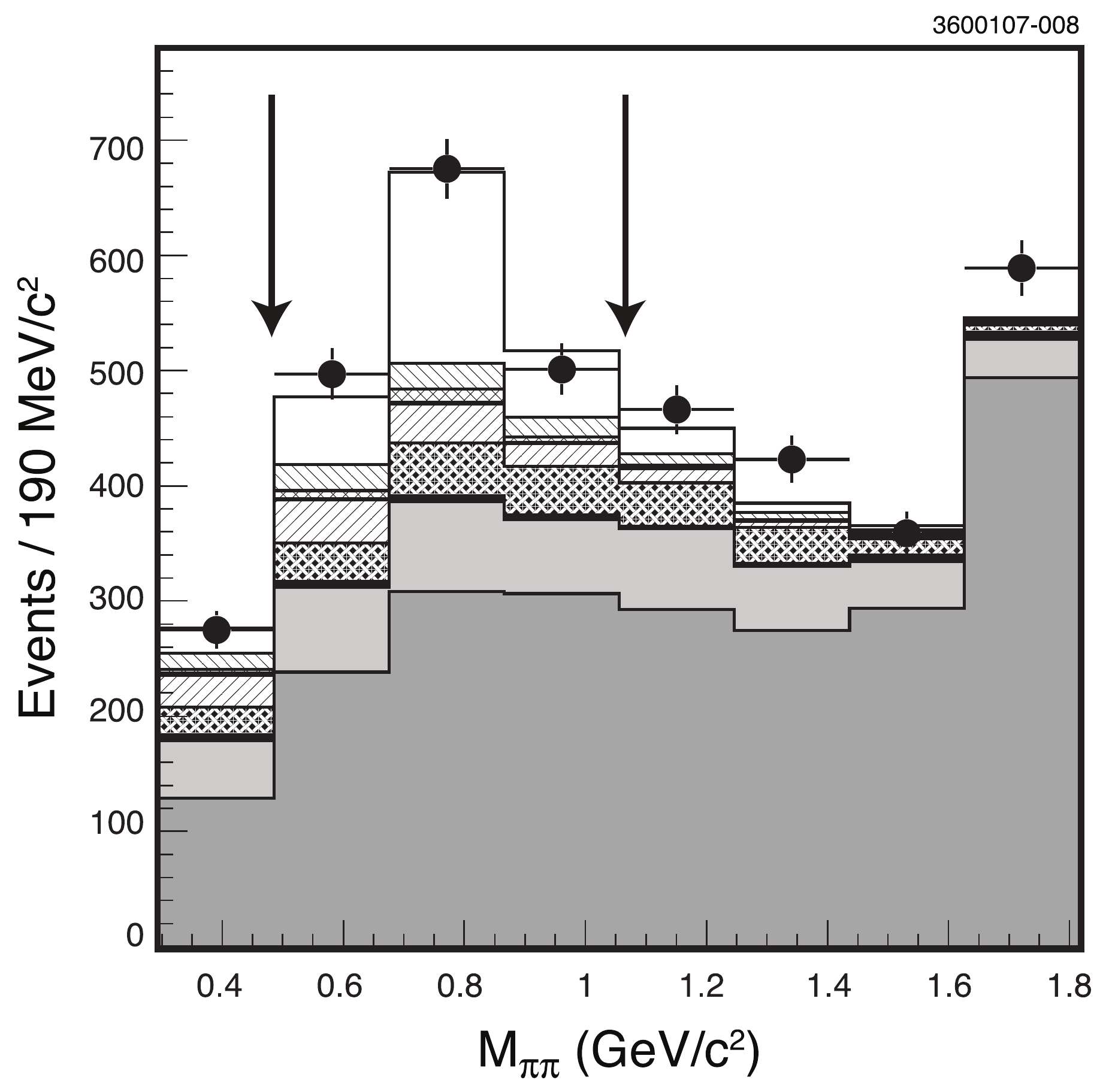}
\caption{$M_{\pi\pi}$ in the restricted signal region for the combined $B\to\rho\ell\nu$ modes, with components as described in Figure~\ref{fig:pifit}. Arrows indicate the range included in the fit. }
\label{fig:mpipi}
\end{figure}

\begin{table}[b]
\caption{Summary of the phase space subregions and the partial and total branching fraction results.}
\label{tab:results}
\begin{tabular}{cccc}\hline\hline
~ & $q^2$ [GeV$^2$]    & $\cos\theta_{W\ell}$ & $\mathcal{B}$ [$10^{-4}$] \\ \hline
$B^0\to\pi^-\ell^+\nu$   & 0 - 2   & $^-1$ - 1  & $0.13\pm 0.07 \pm 0.02$ \\ 
~                           & 2 - 8   &$^-1$ - 1 & $0.27\pm 0.08 \pm 0.03$ \\
~                           & 8 - 16 & $^-1$ - 1 & $0.56\pm 0.09 \pm 0.05$ \\
~                           & $>16$ & $^-1$ - 1 & $0.41\pm 0.08 \pm 0.04$ \\ 
~                           & \multicolumn{2}{c}{all phase space} & ${\bf 1.37 \pm 0.15 \pm 0.11}$ \\
$B^0\to\rho^-\ell^+\nu$ & 0 - 2   & $^-1$ - 1 & $0.45\pm 0.20\pm 0.15$ \\
~                             & 2 - 8   & $^-1$ - 1 & $0.96\pm 0.20\pm 0.29$ \\
~                             & 8 - 16 &  0 - 1 & $0.75\pm 0.16\pm 0.14$ \\
~                             & $>16$    &  0 - 1 & $0.35\pm 0.07\pm 0.05$ \\
~                             & $>8$      & $^-1$ - 0 & $0.42 \pm 0.18 \pm 0.31$ \\ 
~                             & \multicolumn{2}{c}{all phase space} & ${\bf 2.93 \pm 0.37 \pm 0.37}$ \\
$B^0\to\eta\ell^+\nu$ & \multicolumn{2}{c}{all phase space} & ${\bf 0.44 \pm 0.23 \pm 0.11}$ \\ 
$B^0\to\eta^\prime\ell^+\nu$ & \multicolumn{2}{c}{all phase space} & ${\bf 2.66 \pm 0.80 \pm  0.56}$ \\ \hline \hline
\end{tabular}
\end{table}

Table~\ref{tab:results} summarizes our branching fraction results.  The fit yielded $-2\ln{\cal L}=541$, and we note that the statistical errors on some of the 532 bins are not Gaussian.
Figure~\ref{fig:pifit} illustrates a $M_{h\ell\nu}$ projection of the data and fit components for the $B\to\pi\ell\nu$ mode.  The signal, peaking at $M_{h\ell\nu}=M_B,$ is clearly visible.  We also show $M_{\pi\pi}$ (Figure~\ref{fig:mpipi}) 
for $B\to\rho\ell\nu$.  In all cases the data are nicely modeled by the fit.

A detailed summary of the systematic errors can be found in~\cite{PRD}.  The dominant experimental systematic error is neutrino reconstruction efficiency.  This uncertainty arises mainly from systematic effects within the MC simulation, and we have estimated it by randomly discarding or smearing the reconstructed tracks and showers based on independent studies of tracking and calorimeter resolution and efficiency.  We also vary the $B\to X_c\ell\nu$ form factors and branching fractions within present experimental limits.  To estimate systematic bias due to non-resonant $B\to\pi\pi\ell\nu$ contributions, we study reconstructed $B\to\pi^0\pi^0\ell\nu$ candidates.  Non-resonant $\pi\pi\ell\nu$ components generated with $\rho$ lineshapes are included in the fit with relative strengths of the various charge combinations constrained by isospin and angular momentum considerations~\cite{Athar:2003yg}.  The $\pi^0\pi^0\ell\nu$ mode then limits the level that the $\rho$ lineshape can be projected out of the (unknown) non-resonant shape.  No statistically significant non-resonant bias of the $B\to\rho\ell\nu$ rate is observed. 

While our fitting approach significantly reduces dependence on {\em a priori} knowledge of the FFs, residual FF dependence due to efficiency variation within each phase-space region remains.   We study this dependence with several other FF calculations for $B\to\pi\ell\nu$~\cite{isgw2,spd,ball04pi} and $B\to\rho\ell\nu$~\cite{meli,isgw2,ukqcd98}, chosen to span conservative ranges in shape.  We vary the $B\to\rho\ell\nu$ FF keeping the $B\to\pi\ell\nu$ FF fixed to its nominal, and {\it vice versa.}  The systematic uncertainty is estimated be one-half of the total spread of the results.

We find evidence for $\mathcal{B}\left(B^+\to\eta^{\prime}\ell^+\nu\right)=(2.66\pm0.80\pm0.56 )\times 10^{-4}$ at $3\sigma$ significance and set an upper limit  at the 90\% confidence level on the branching fraction for the decay $\mathcal{B}\left(B^+\to\eta\ell^+\nu\right) < 1.01\times10^{-4}$.  The latter results require $\mathcal{B}\left(B^+\to\eta^{\prime}\ell^+\nu\right) / \mathcal{B}\left(B^+\to\eta\ell^+\nu\right) > 2.5$ at 90\% confidence level, an indication of enhanced coupling to the singlet component of the $\eta^\prime$.  From a model dependent study (see \cite{PRD}) of the $\eta^{(')}$ rates following Beneke and Neubert~\cite{etap:Neubert2003}, we estimate ${\cal B}(B^-\to K^-\eta^{\prime})=(84^{+35}_{-25}{}^{+53}_{-24})\times~10^{-6}$, in good agreement with the current experimental value of ${\cal B}(B^-\to K^-\eta^{\prime})=(71\pm4)\times 10^{-6}$ \cite{pdg06}.  The $3\sigma$ significance of our $\mathcal{B}\left(B^+\to\eta^{\prime}\ell^+\nu\right)$ result is estimated by using a toy MC procedure to combine systematic and statistical errors and determine the probability that zero rate would produce a result at or above our measured value~\cite{PRD}.

Our $B\to\rho\ell\nu$ results currently provide the most precise test of the $B\to\rho$ form factor calculations to date. We fit the predicted form factor shape to the measured rate in the five phase space regions.  The resulting fit has four degrees of freedom, and we find the $\chi^2$ to be 4.5, 9.0, and 4.3, for the LCSR~\cite{ball04rho}, ISGW2~\cite{isgw2}, and UKQCD~\cite{ukqcd98} calculations.  The LCSR and LQCD calculations, extrapolated over all phase space, agree with the data much better than the ISGW2 model.

The full potential for extraction of $|V_{ub}|$ will only be realized once techniques that can utilize a broad range of $q^2$ have matured.  For now, we extract a value for $|V_{ub}|$ by combining our measured rate for $B^0\to\pi^-\ell^+\nu$ in the $q^2 > 16~\text{GeV}^2$  region with the $B^0$ lifetime and the partial width prediction, $\Gamma/|V_{ub}|^2$, from LQCD~\cite{Gulez:2006dt}, which calculates FFs using a minimal number of assumptions.
  We find $|V_{ub}| = \left(3.6\pm 0.4 \pm 0.2 {}^{+0.6}_{-0.4}\right)\times 10^{-3}$.  The errors are, in order, statistical, experimental systematic, and theoretical. 

In summary we have measured the branching fractions $\mathcal{B}\left(B^0\to\pi^-\ell^+\nu\right)=(1.37 \pm 0.15 \pm 0.11 )\times 10^{-4}$, $\mathcal{B}\left(B^0\to\rho^-\ell^+\nu\right) = (2.93 \pm 0.37 \pm 0.37 )\times 10^{-4}$, and $\mathcal{B}\left(B^+\to\eta^{\prime}\ell^+\nu\right)=(2.66\pm0.80\pm0.56 )\times 10^{-4}$.  We find $|V_{ub}| = \left(3.6\pm 0.4 \pm 0.2 {}^{+0.6}_{-0.4}\right)\times 10^{-3}$ from $\mathcal{B}\left(B^0\to\pi^-\ell^+\nu\right)$ in the $q^2>16~\text{GeV}^2$ region, where our results are consistent with other measurements~\cite{bb:BABARy2007,bb:BABARy2006,bb:BELLEy2007}.  These results constitute one of the two most precise measurements of $\mathcal{B}\left(B^0\to\pi^-\ell^+\nu\right)$~\cite{bb:BABARy2007}, the most precise measurement of $\mathcal{B}\left(B^0\to\rho^-\ell^+\nu\right)$, and the first evidence for $B\to\eta^\prime\ell\nu$.  

We gratefully acknowledge the effort of the CESR staff in providing us with excellent luminosity and running conditions. This work was supported by the A.P.~Sloan Foundation, the National Science Foundation, the U.S. Department of Energy, and the Natural Sciences and Engineering Research Council of Canada.

\end{document}